\documentclass[secnumarabic,showpacs,amsmath,amssymb,floatfix,superscriptaddress,nofootinbib, nobibnotes, aps, prl, twocolumn]{revtex4-1}
\usepackage{graphicx}
\usepackage{epstopdf}
\usepackage{color}         			
\usepackage[normalem]{ulem}			
\usepackage{soul} 						 	
\usepackage{lineno}
\usepackage{multibib}

\usepackage{newfloat}

\DeclareFloatingEnvironment[fileext=lop]{Extended Data Figure}

\begin{document}


\title{Creation of a Chiral Bobber Lattice in Helimagnet-Multilayer Heterostructures}

\author{Kejing Ran$^\ast$}
\affiliation{School of Physical Science and Technology, ShanghaiTech University, Shanghai 200031, China}
\affiliation{ShanghaiTech Laboratory for Topological Physics, ShanghaiTech University, Shanghai 200031, China}

\author{Yizhou Liu$^\ast$}
\affiliation{RIKEN Center for Emergent Matter Science (CEMS), Wako 351-0198, Japan}

\author{Yao Guang}
\affiliation{Beijing National Laboratory for Condensed Matter Physics, Institute of Physics, Chinese Academy of Sciences, Beijing 100190, China}

\author{David M. Burn}
\affiliation{Diamond Light Source, Harwell Science and Innovation Campus, Didcot, Oxfordshire, OX11 0DE, UK}



\author{Gerrit \surname{van~der~Laan}}
\affiliation{Diamond Light Source, Harwell Science and Innovation Campus, Didcot, Oxfordshire, OX11 0DE, UK}

\author{Thorsten Hesjedal}
\affiliation{Clarendon Laboratory, Department of Physics, University of Oxford, Parks Road, Oxford, OX1~3PU, United Kingdom}

\author{Haifeng Du}
\affiliation{The Anhui Key Laboratory of Condensed Matter Physics at Extreme Conditions, High Magnetic Field Laboratory and University of Science and Technology of China, Chinese Academy of Science (CAS), Hefei, Anhui 230031, China.}

\author{Guoqiang Yu}
\affiliation{Beijing National Laboratory for Condensed Matter Physics, Institute of Physics, Chinese Academy of Sciences, Beijing 100190, China}

\author{Shilei Zhang }
\affiliation{School of Physical Science and Technology, ShanghaiTech University, Shanghai 200031, China}
\affiliation{ShanghaiTech Laboratory for Topological Physics, ShanghaiTech University, Shanghai 200031, China}

\date{\today}

\begin{abstract}
A chiral bobber is a localized three-dimensional magnetization configuration, terminated by a singularity.
Chiral bobbers coexist with magnetic skyrmions in chiral magnets, lending themselves to new types of skyrmion-complementary bits of information.
However, the on-demand creation of bobbers, as well as their direct observation remained elusive.
Here, we introduce a new mechanism for creating a stable chiral bobber lattice state via the proximity of two skyrmion species with comparable size.
This effect is experimentally demonstrated in a Cu$_2$OSeO$_3$/[Ta/CoFeB/MgO]$_4$ heterostructure in which an exotic bobber lattice state emerges in the phase diagram of Cu$_2$OSeO$_3$.
To unambiguously reveal the existence of the chiral bobber lattice state, we have developed a novel characterization technique, magnetic truncation rod analysis, which is based on resonant elastic x-ray scattering.
\end{abstract}


\maketitle

Magnetic skyrmions are two-dimensional, particle-like solitonic field configurations with extraordinary topological properties \cite{BogdanovJETP:89,TokuraNatnano:13}.
Although most magnetic systems in which skyrmions exist are essentially three-dimensional (3D) objects, the skyrmion is classified by elements of the homotopy group $\pi_3(S^2)$ \cite{Mermin_topology_review_RMP_79}.
Recently, complex 3D magnetic textures were found which are classified by the $\pi_2(S^3)$ group \cite{Bogdanov_surface_twist_PRB_13,  Rybakov_chiral_bobber_PRL_15, Bogdanov_surface_twist_PRL_16, Ivan_hopfions_lq_Natmater_16, Leonov_3D_Tube_APL_16, Leonov_3D_Tube_IOP_16,  DER_skyrmion_18,Leonov_CSO_LTEM_3DTube:2018, Du_FeGe_EH_attractive_PRL_18, Du_FeGe_EH_bobber_Natnano_18, Beni_Bobber_PRL_18, Ohio_FeGe_LTEM_bobber_PRM_18, Leonov_bobber_toron_PRB_18, Yizhou_Hopfion_PRB_18, Kiselev_bobber_pair_PRB_20, Fischer_review_3D_nanostructure_APLM_20}, providing an ideal playground for studying topological defects and magnetic monopole-related science. Therefore, the experimental exploration of 3D magnetic structures with topological properties has become an important task.

A major class of materials that hosts 3D skyrmion structures are chiral magnets, such as MnSi \cite{Pf_MnSi_Science_09},
FeCoSi \cite{Tokura_FeCoSi_LTEM_Nature_10},
FeGe \cite{Tokura_FeGe_LTEM_Natmater_11},
Cu$_2$OSeO$_3$ \cite{Tokura_CuOSeO_LTEM_Science_12},
CoZnMn \cite{Tokura_CoZnMn_single_crystal_Natcomm_15}, and so on.
In these bulk materials (with periodic boundary condition), a particular energy hierarchy leads to the formation of multidimensional solitons:
$w = A(\nabla \mathbf{m}^2) + D \mathbf{m} \cdot (\nabla \times \mathbf{m}) - \mathbf{B}\cdot\mathbf{m} + w_D$,
where $\mathbf{m}(x,y,z)$ is the real-space magnetization configuration. The energy density $w$ contains first three local terms, i.e., exchange interaction with stiffness constant $A$, Dzyaloshinskii-Moriya interaction (DMI) with a strength of $D$, a Zeeman term that scales with external field $\mathbf{B}$, and a non-local dipole-dipole interaction term $w_D$.
The system is thus described by a modulated structure, with characteristic periodicity $\lambda_\mathrm{h} = 4\pi A/D$. It was recognized that under certain perturbations, such as magnetocrystalline anisotropy or thermal fluctuations, a skyrmion lattice state can form in a narrow region of the temperature-magnetic field phase diagram between the lower and upper critical fields, $B_{\textnormal{A1}}$ and $B_{\rm{A2}}$, respectively, close to the transition temperature, although the lowest energy ground state is the one-dimensional modulated conical state \cite{Bogdanov_JMMM_94, Pf_MnSi_Science_09, TokuraNatnano:13, Monchesky_Wilson_noskyrmion_PRB_14, Pf_Cu2OSeO3_LT_SkX_Natphys_18}.
Under these conditions, the 2D skyrmion lattices are stacked along the field direction, forming the skyrmion tube lattice (SkTL) structure.

For chiral magnets with a finite thickness that is comparable to $\lambda_\mathrm{h}$, the terminating surfaces break translational symmetry, leading to surface twist effect \cite{Bogdanov_surface_twist_PRB_13, Monchesky_MnSi_surface_twist_PRB_14, Bogdanov_surface_twist_PRL_16, DER_skyrmion_18}.
Surface twisting is responsible for a number of phenomena:\
(i) a modulation of the 3D skyrmion tubes along the field direction (with varying helicity angle) \cite{Bogdanov_surface_twist_PRL_16, DER_skyrmion_18, Depth_profile_18}, and consequently,
(ii) an enhanced energetic stability over the conical phase, witnessed by the largely expanded skyrmion region in the phase diagrams \cite{Tokura_FeGe_LTEM_Natmater_11,Bogdanov_surface_twist_PRL_16}.
(iii) At higher fields above $B_{\textnormal{A2}}$, a metastable surface state, i.e., the chiral bobbers (ChBs), which represent a new type of 3D topological texture, can be generated through a non-equilibrium process \cite{Rybakov_chiral_bobber_PRL_15, Du_FeGe_EH_bobber_Natnano_18}.

\begin{figure}[ht]
\includegraphics[width=8.6cm]{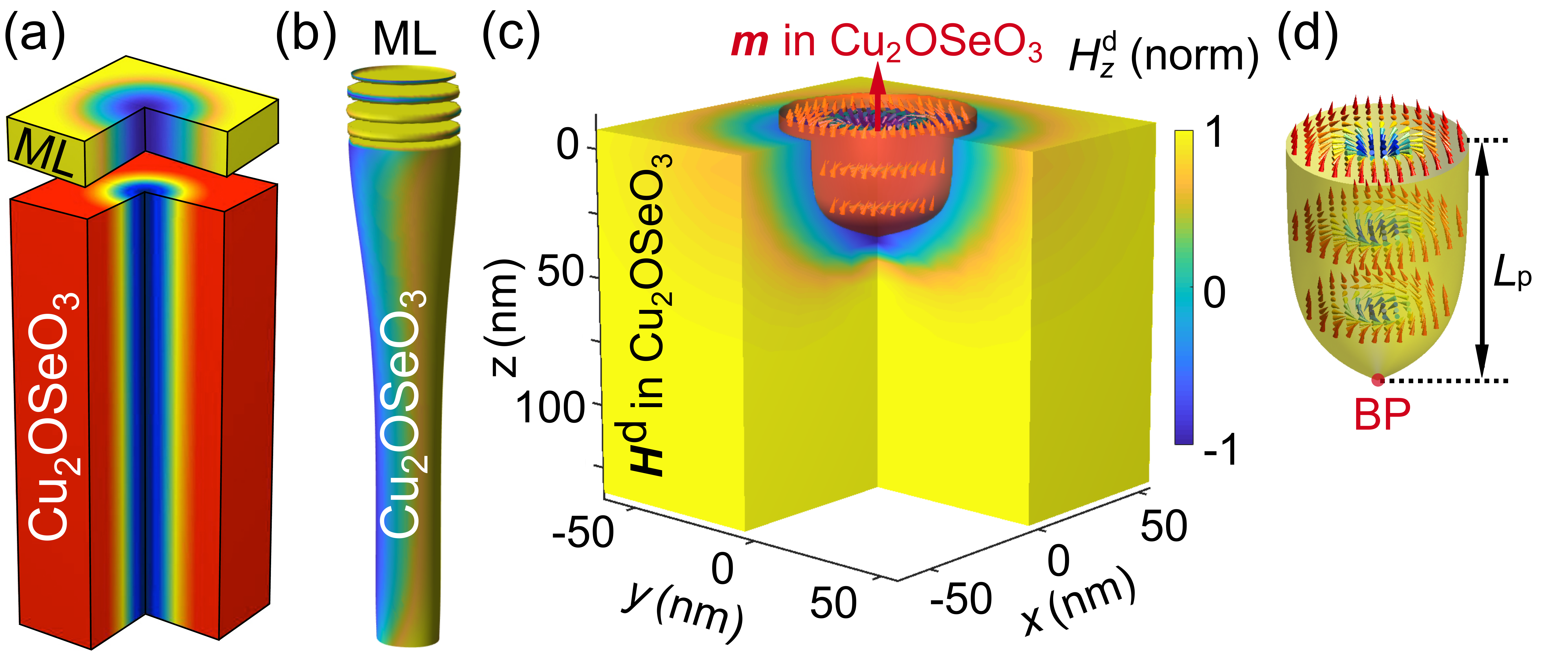}
\caption{ Creation of a chiral bobber structure via heterostructure engineering.
	(a) Illustration of the heterostructure by contacting two different skyrmion species with comparable lateral dimensions.
	(b) Simulation results of the skyrmion tube structure at field between $B_{\textnormal{A1}}$ and $B_{\textnormal{A2}}$.
	(c) $z$-component of the dipolar field distribution from a skyrmion in the multilayer.
	(d) Calculated chiral bobber structure formed in Cu$_2$OSeO$_3$ at a field between the first and second upper critical fields $B_{\textnormal{A2}}$ and $B_{\textnormal{A3}}$, respectively.}
\label{fig_1}
\end{figure}

Chiral bobbers are characterized by a distinct 3D structure, resembling a floating bobber, that consists of a skyrmion stack with continuously reducing diameter away from the surface with a bobber length $L_\textnormal{p}$, eventually shrinking down to a singularity, the Bloch point (BP) \cite{Rybakov_chiral_bobber_PRL_15}.
Such a topological point defect has finite energy, revealing itself as a metastable state at fields larger than $B_{\textnormal{A2}}$, where the conical phase dominates \cite{Rybakov_chiral_bobber_PRL_15, Du_FeGe_EH_bobber_Natnano_18}.
The study of ChBs is of great interest due the presence of the BP, which resembles a monopole structure with emergent dynamical behavior \cite{Rocsh_FeCoSi_monopole, Pf_FeCoSi_LTEM_SciAdv_17, Leonov_bobber_toron_PRB_18, Kiselev_bobber_pair_PRB_20}.
Therefore, the ChB phase provides an excellent platform for the general study of topological defect-related physics.
Moreover, due to the similar energy barriers between ChBs and SkTs, bobbers can serve as promising information carriers for advanced memory devices in conjunction with skyrmions, where the presence of either represents a different information state \cite{Du_FeGe_EH_bobber_Natnano_18, Kiselev_bobber_transport_PRB_19}.

Nevertheless, the creation of bobbers has been a challenging task. So far, the mechanism that stabilizes ChBs largely relies on their metastable nature, i.e., they can be produced by either field-cooling or field-tilting protocols \cite{Du_FeGe_EH_bobber_Natnano_18}.
In both scenarios, bobbers seem to randomly nucleate at arbitrary positions within the surface \cite{Du_FeGe_EH_bobber_Natnano_18}. This prevents further experimental studies of their novel properties, and hinders their use in future racetrack memory.
Recent work hints at the possibility that the exotic interface Rashba DMI can stabilize ChBs in epitaxially grown FeGe thin films \cite{Ohio_FeGe_LTEM_bobber_PRM_18}, however, the unambiguous experimental characterization of the bobber structure, including their shape and penetration length, still remains elusive.
In this Letter, we present a new mechanism to create ChBs in a controlled manner, as well as a new experimental technique for their direct observation.

Figure~\ref{fig_1}(a) shows the key concept of the new mechanism that we adopted to stabilize the chiral bobber lattice (ChBL) in a heterostructure formed by a bulk Cu$_2$OSeO$_3$ crystal and a multilayered (ML) thin film.  Cu$_2$OSeO$_3$ is a well-known SkTL-hosting chiral magnet, which has a helical wavelength $\lambda_\mathrm{h} \approx 56$~nm, a transition temperature $T_\mathrm{C}\approx 57$~K, and an upper critical field $B_{\rm{A2}}\approx 37$~mT \cite{Cu2OSeO3_REXS_NL_2016}.
Using resonant elastic x-ray scattering (REXS), it was shown that surface twisting is a pronounced effect at the Cu$_2$OSeO$_3$ $[001]$ surface. Nevertheless, a bobber phase has not been observed in chiral bulk magnets \cite{Cu2OSeO3_REXS_NL_2016, DER_skyrmion_18}.
Our strategy for obtaining bobbers is to make use of proximity coupling between two skyrmion systems with comparable lateral dimensions.
In case of Cu$_2$OSeO$_3$ bulk crystals, we selected a tunable [Ta/CoFeB/MgO]$_n$ multilayer structure.
In an earlier study, we have shown that skyrmions with diameters of $\sim$100~nm can be stabilized in such ML systems in a relatively broad range of magnetic fields and temperatures, covering the range of $B_{\rm{A1}}$ to $B_{\rm{A2}}$ of Cu$_2$OSeO$_3$ \cite{N4_REXS_19}.
Consequently, two interactions will be dominating the heterostructure. First, the interface between Cu$_2$OSeO$_3$ and CoFeB can be mediated by a thin Ta layer, supporting a Ruderman-Kittel-Kasuya-Yosida (RKKY)-like exchange interaction $J_{\text{RKKY}}$ across the two interface layers $i$ and $j$.

Such an interfacial exchange locks the positions of the skyrmions in the ML, and aligns them with the bulk skyrmions, 
as shown by our micromagnetic simulations using \texttt{MuMax3} in Fig.~\ref{fig_1}(b) \cite{MuMax}.
The interfacial exchange effect connects the two SkT species, and accommodates their difference in lateral dimension by forming a smoothly varying 3D funnel-like structure.
Second, due to the larger saturation magnetization $M_\text{S}$ in the ML, the dipole-dipole interaction leads to a relatively strong non-local stray field, influencing the SkTs in the bulk crystal.

Figure~\ref{fig_1}(c) shows the calculated distribution of the $z$-component of the dipolar field $H_z^\mathrm{d}$ in the bulk region. The shape of the stray field and the energy density $w_D$ resembles that of a bobber-like structure. For the field range of $B_\textnormal{A1}$$<$$B$$<$$B_\textnormal{A2}$, the intrinsic interactions coming from bulk Cu$_2$OSeO$_3$ dominate over the dipolar interaction from the multilayer, forming a standard SkTL phase---unaffected by interfacial effects.
At fields slightly above $B_{\rm{A2}}$, however, the SkTL state in Cu$_2$OSeO$_3$ starts to evolve into the conical phase, following a first-order type phase transition \cite{Pf_MnSi_fluctuation_1st_transition_PRB_13}.
Nevertheless, the skyrmion state in the ML remains intact, taking on a funnel-like structure in the near-surface region of Cu$_2$OSeO$_3$. Subsequently, the ML skyrmion dipolar field [Fig.~\ref{fig_1}(c)] breaks the SkTL in the bulk, leading to the formation of bobbers with a finite  depth  $L_\mathrm{p}$, as shown in Fig.~\ref{fig_1}(d).

\begin{figure}[h]
\includegraphics[trim = 0cm 0cm 0cm 0cm, clip=true, width=9cm]{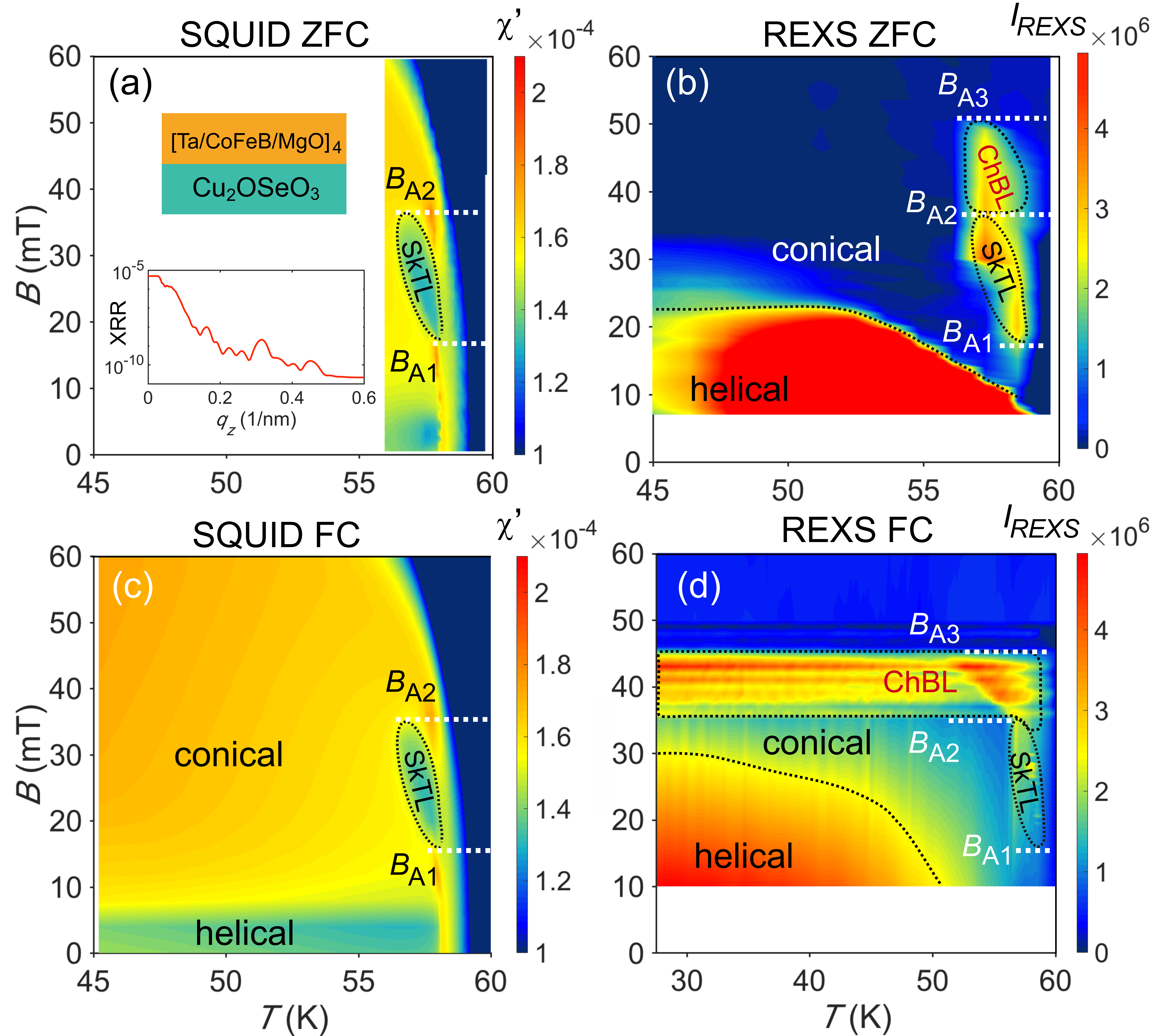}
\caption{Phase diagrams mapped by (a),(c) ac susceptibility and (b),(d) REXS on the same ML/Cu$_2$SeO$_3$ sample.
	(a),(b) Measurements carried out after zero-field-cooling and (c),(d) after field-cooling.
	The lower critical field $B_\textnormal{A1}$ and the upper critical field $B_\textnormal{A2}$ enclose the SkTL pocket, whereas the ChBL state reached from $B_\textnormal{A2}$ to the second upper critical field $B_\textnormal{A3}$.
}
\label{fig_2}
\end{figure}

The [Ta/CoFeB/MgO]$_4$ multilayer structure was grown by magnetron sputtering on a finely-polished Cu$_2$OSeO$_3$ [001] substrate, following the recipe in Ref.\ \cite{N4_REXS_19}.
The inset in Fig.~\ref{fig_2}(a) shows the x-ray reflectivity (XRR) profile, characterized by a smoothly decaying specular intensity with superimposed fringes, suggesting a well-defined heterostructure interface and excellent superstructure film quality.
Next, the magnetic phase diagram of the \emph{same} heterostructure sample was mapped out using two complementary techniques:
(i) The ac susceptibility, measured in a superconducting quantum interference device (SQUID) magnetometer, is dominated by the Cu$_2$OSeO$_3$ \emph{bulk} properties.
(ii) REXS in reflection geometry, with the photon energy tuned to Cu $L_3$ edge, measures the \emph{surface} properties of Cu$_2$OSeO$_3$ only, down to a depth of $<$100\,nm \cite{Depth_profile_18}.

Figure~\ref{fig_2}(a) and (b) show the phase diagrams mapped by ac susceptibility and REXS for $\mathbf{B} \parallel [001]$ and using zero-field cooling.
The system exhibits the typical phase diagrams for skyrmion-hosting chiral magnets, as shown in Fig.~\ref{fig_2}(a), from which the SkTL phase can be clearly singled out near $T_\mathrm{C}$.
The lower and upper critical field values of $B_\textnormal{A1}$ and $B_\textnormal{A2}$ are consistent with other reports on Cu$_2$OSeO$_3$ \cite{Cu2OSeO3_REXS_NL_2016}.
On the other hand, REXS is able to unambiguously confirm the SkTL phase with its characteristic six-fold-symmetric diffraction pattern \cite{Cu2OSeO3_REXS_NL_2016, Cu2OSeO3_Domain_imaging_APL_16, Winding_Natcommun_17}.
As shown in Fig.~\ref{fig_2}(b), we surprisingly find an additional phase pocket, located above the critical field $B_\textnormal{A2}$.
This phase again shows a six-fold-symmetric REXS pattern, identical to the one of the SkTL lattice \cite{SM}.
The pattern even survives in fields above 50\,mT (at 57\,K), nevertheless, this phase is invisible to bulk-sensitive SQUID measurements.
It is worth noting that throughout our REXS experiments, the photon energy was tuned to the Cu $L_3$ edge (931.25\,eV), which exclusively probes the surface region of Cu$_2$OSeO$_3$ \cite{Gerrit_REXS_Physique_08}.
We thus ascribe the origin of the phase pocket region from $B_\textnormal{A2}$ to the second upper critical field $B_\textnormal{A3}$ to the interfacial effect from the multilayer [labeled \emph{chiral bobber lattice} (ChBL) state in the figure for reasons that will soon become clear].

Figures~\ref{fig_2}(c) and \ref{fig_2}(d) show the phase diagrams after field-cooling from above $T_\mathrm{C}$. It is recognized that a metastable skyrmion phase can be observed at temperatures far below $T_\mathrm{C}$, which is obtained by field-cooling as well \cite{Rocsh_FeCoSi_monopole}, however, which reveals itself in \emph{bulk} ac susceptibility measurement \cite{Pf_MnSi_ac_SQUID}.
In our case, no metastable skyrmion state was in the SQUID measurements shown in Fig.~\ref{fig_2}(c).
Instead, the phase diagram is almost identical to the one shown in Fig.\ 2(a).
The standard SkTL phase, which exists between $B_{\rm{A1}}$ and $B_{\rm{A2}}$, can also be easily singled out in REXS measurements as shown in Fig.~\ref{fig_2}(d).
More interestingly, the ChBL phase is developing metastable behavior when cooling down the system, while the critical field values of $B_{\rm{A2}}$ and $B_{\rm{A3}}$ remain roughly consistent with those obtained from Fig.~\ref{fig_2}(b).
This additional high-field phase, and its metastable behavior, provides strong clues hinting at the possible existence of chiral bobbers.

In order to fully characterize the magnetic structure of the ChBL phase, we developed magnetic truncation rod (MTR) analysis, a new REXS-based characterization technique.
In general, crystalline truncation rods occur in many surface diffraction processes in which the incidence waves (e.g., x-rays or electrons) are sensitive to the terminating surface of a crystal, either due to a small incident angle or shallow probing depth \cite{RobinsonCTRPRB:86}.
In such a scenario, delta-function-like diffraction peaks in reciprocal space are extending into rods in the direction of the surface normal. By analyzing the rod profile, one is able to `reconstruct' the detailed near-surface structure, which is especially useful when the structure shows a depth dependence \cite{RobinsonCTRPRB:86}.

\begin{figure}[h]
\includegraphics[trim = 0cm 0cm 0cm 0cm, clip=true, width=7.5cm]{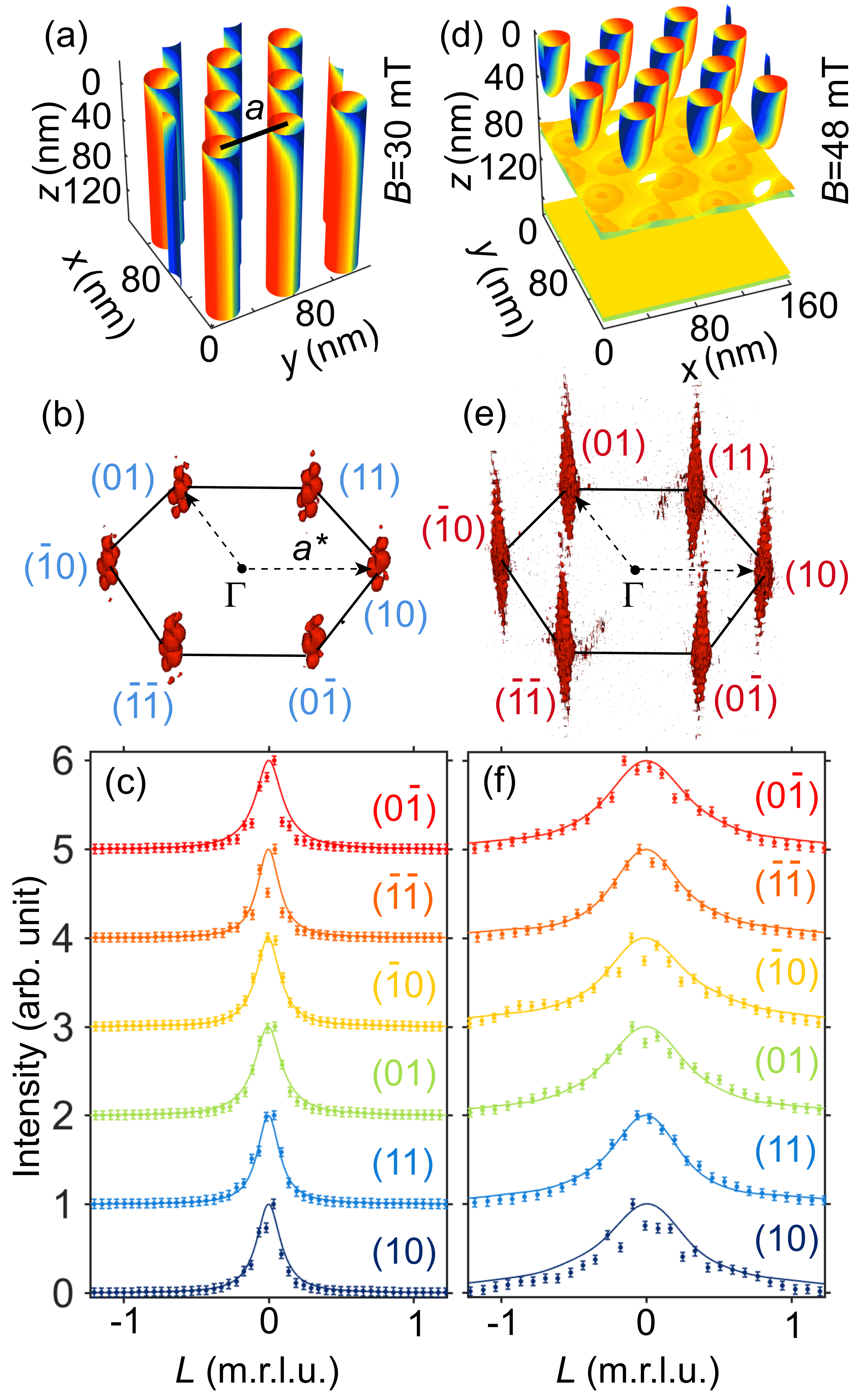}
\caption{Magnetic truncation rod analysis of skyrmions and chiral bobbers.
	(a) Micromagnetic simulation results of the SkTL phase. The colors of the $m_z=0$ isosurface illustrate the local $m_x$ and $m_y$ in-plane magnetization components.
	(b) REXS intensity distribution in 3D reciprocal space, measured in the SkTL state at 57\,K and 30\,mT.
	(c) MTR profiles for six rods in the SkTL state as indicated. Circles represent experimental data (with error bars), and solid lines are fitted truncation roads using the simulated magnetic structure shown in (a).
	(d)-(f) Corresponding micromagnetic simulation results, REXS pattern, and MTR profiles for the chiral bobber lattice measured at 57\,K and 48\,mT.}
\label{fig_3}
\end{figure}

Here, we extend the crystalline truncation rod theory to magnetic structures probed by soft x-ray resonant magnetic diffraction \cite{SM}.
The SkTL phase in Cu$_2$OSeO$_3$ can be regarded as long-range-ordered magnetic crystal, shown in Fig.~\ref{fig_3}(a).
The hexagonal unit cell has a (magnetic) lattice constant of $a=65$\,nm, with the motif being a single skyrmion.
Such a two-dimensional skyrmion crystal is associated with the reciprocal space pattern shown in Fig.~\ref{fig_3}(b) with six 2D lattice peaks, i.e., (10), (11), (01), ($\bar 1$0), ($\bar 1$$\bar 1$), and (0$\bar 1$).
The origin is at the $\Gamma$ point and the reciprocal lattice constant is $a^\ast = 4\pi/\sqrt3a$. Likewise, the magnetic Miller indices ($H$,$K$,$L$) can be defined in terms of \emph{magnetic reciprocal lattice units} (m.r.l.u.). For soft x-rays at resonance (with the Cu $L_3$ edge in this case), the penetration length $\Lambda$ is usually below 100\,nm \cite{Gerrit_REXS_Physique_08}, suggesting a pronounced effect of surface magnetic diffraction.
Consequently, the six peaks are extended into rods along $L$, i.e., \emph{magnetic} truncation rods. They are thus expressed by a structure factor of the form
$\mathbf{F}(\mathbf{q}) = -\hat{M} \, [ iF^{(1)}(\epsilon_s^\ast \times \epsilon_i)\cdot \mathbf{\hat{n}}]$,
%
where $\mathbf{q}$ is the elastic scattering momentum transfer, which has to equate a reciprocal lattice vector $(H,K,L)$ (of the magnetic lattice) to satisfy the diffraction condition.
$F^{(1)}$ relates to the energy dependent dipole-transition amplitude of the magnetic scattering, $\mathbf{\epsilon}_s$ and $\mathbf{\epsilon}_i$ denote the unit vectors of the scattered and incident x-ray polarization, and $\mathbf{\hat{n}} = (n_1, n_2, n_3)$ is the unit vector of the local magnetic moment.

The quantity $\hat{M}$ in the structure factor $\mathbf{F}(\mathbf{q})$ can be expressed as
\begin{equation}
\begin{split}
\hat{M} = \frac{1}{\Lambda} \sum_{z=0}^{\infty}e^{-2z \sec\alpha / \Lambda}  \sum_{n} e^{2\pi i\mathbf{q\cdot r}_n}   \,\,\,,
\label{eq_11}
\end{split}
\end{equation}

\noindent where $z=0$ corresponds to the top surface, $\alpha$ is the incident angle, and the summation is carried out over all atomic positions $n$ in the plane. For the photon energy tuned to the Cu $L_3$ edge, the x-ray attenuation length $\Lambda = 93.9$\,nm \cite{Depth_profile_18}.
Equation (1) allows us to calculate the MTRs in a layer-by-layer fashion. For $\mathbf{\epsilon}_s$ summed over both polarizations, the MTR intensity $I(\mathbf{q})=|\mathbf{F(q)}|^2$ \cite{Winding_Natcommun_17} can be written as
$I(\mathbf{q}) = \text{tr} [ \hat{M}f_m u_i \hat{M}f_m^\dagger]$ (for details see Supplemental Material \cite{SM}).


%



Figure~\ref{fig_3}(b) shows the MTR data for Cu$_2$OSeO$_3$/ [Ta/CoFeB/MgO]$_4$ measured in the SkTL phase at 57\,K and 30\,mT. The total intensity is the sum of the intensities measured with left- and right-circularly polarized incident light.
The six magnetic Bragg peaks have a confined blob-like shape, which is slightly extended along the $L$ direction.
We ascribe this peak structure to the natural broadening owing to the finite penetration depth of soft x-rays, reflecting the bulk character of 3D skyrmion tubes.
The detailed rod profiles along $L$ are shown in Fig.~\ref{fig_3}(c).
The experimental data for the six rods can be almost perfectly reproduced in simulations of the MTRs based on the equation for $I(\mathbf{q})$ and using the micromagnetic simulation results shown in Fig.\ \ref{fig_3}(a).
It is worth mentioning that we also measured MTRs in the SkTL phase of a pristine Cu$_2$OSeO$_3$ $[001]$ crystal (without a ML), which shows the same line shapes cutting through the rods along $L$.

Next, we measured MTRs in the bobber phase at 57\,K and 48\,mT.
The 3D plot of the rod intensities is shown in Fig.~\ref{fig_3}(e).
Although the ChBL state has the same reciprocal lattice in the $H$-$K$ plane for $L=0$, the rod profiles are significantly elongated along $L$ --- a signature of surface diffraction --- with the x-rays probing the shallow magnetization pattern buried right underneath the surface.
Comparing the contrast in Figs.~\ref{fig_3}(b) and (e) suggests a very different depth profile of skyrmions and bobbers near the very interface of the heterostructure.

Figure~\ref{fig_3}(f) shows the MTR profiles of the six rods along $L$ in the bobber phase. The peaks are significantly broadened compared to the skyrmion profiles shown in Fig.~\ref{fig_3}(c).
By performing systematic numerical MTR simulations of micromagnetic models, we find that the contrast is sensitive to the 3D shape of the bobbers, as well as their extension in depth, $L_{\rm{p}}$ \cite{SM}.
The six rods all have slightly different shapes as a result of a geometrical effect, i.e., as they are distributed at different azimuthal angles within the $x$-$y$ plane, the x-rays `see' them from different perspective.
Nevertheless, one should be able to fit all six rods at the same time using one 3D magnetic structure model. The solid lines in Fig.\ \ref{fig_3}(f) are fitted rod profiles using the ChBL model obtained from micromagnetic simulations, shown in Fig.\ \ref{fig_3}(d).
The bobbers in Cu$_2$OSeO$_3$ form a well-ordered lattice with a bobber depth of $L_{\rm{p}} = 40$\,nm.
The bobber shape, described by the skyrmion diameter evolution along $z$, can be extracted from our analysis.
It is important to point out that such precise $L_{\rm{p}}$, as well as bobber shape with BP at the bottom are highly restricted by our numerical refinement, while other possible 3D near-surface structures can be excluded \cite{SM}.
The experimental data agrees well with our micromagnetic model, providing unambiguous evidence for the existence of a chiral bobber lattice.

It is worth mentioning that the rod broadening starts to take place just above $B_{\rm{A2}}$, while we did not observe a gradual change of the peak width upon increasing the field. This indicates that the transition between SkTL and ChBL is accompanied by a sudden change of $L_{\rm{p}}$.
The value of $\sim$40\,nm is much larger than that of metastable bobbers in confined thin plate geometries, in which one can expect $L_{\rm{p}}$ to be smaller than $\lambda_\mathrm{h}/2 = 28$~nm \cite{Rybakov_chiral_bobber_PRL_15}. This difference in $L_{\rm{p}}$ values is due to the different stabilization mechanisms: whereas the bobber lattice state is stabilized via the interactions across the interface, geometrically confined geometry bobbers are due to the surface twist effect.
Note that we also performed control experiments on Cu$_2$OSeO$_3$/Ta and Cu$_2$OSeO$_3$/Pt samples \cite{SM}, which excludes possible contributions of Rashba spin-orbit coupling and induced surface anisotropy \cite{LeePRL:2020} to the formation of ChBL.
Furthermore, everywhere within the ChBL phase between $B_{\rm{A2}}$ and $B_{\rm{A3}}$, $L_{\rm{p}}$ maintains a value of $(40 \pm 5)$\,nm. It is expected that by tuning the materials parameters of the ML, such as the CoFeB  thickness and the repetition number $n$, $L_{\rm{p}}$ can be effectively engineered.

In summary, we uncovered a new mechanism for creating long-range-ordered chiral bobber lattices by coupling two skyrmion lattice states across a chiral bulk crystal-ferromagnetic heterostructure interface.
Chiral bobbers are attractive topological structures, which remained elusive as their controlled stabilization was challenging.
Our coupling approach unlocks a wide range of opportunities for the detailed study of the physical properties of bobbers, such as transport and dynamics.
Further, magnetic truncation rod analysis in REXS is a powerful technique for studying skyrmions and bobbers, for determining complex 3D magnetic structures in general.
The controlled nucleation of ChBs, which we have demonstrated here, is the prerequisite for their use in skyrmion-bobber memory \cite{Du_FeGe_EH_bobber_Natnano_18, Kiselev_bobber_transport_PRB_19}.

{\it{Acknowledgements--}} The REXS experiments were carried out on beamline I10 at the Diamond Light Source, U.K., under proposal SI20182. The authors thank the Analytical Instrumentation Center (SPST-AIC10112914), the School of Physical Science and Technology (SPST), ShanghaiTech University for SQUID and XRD characterizations. S.L.Z.\ acknowledges the starting grant from ShanghaiTech University and 
the National Key Research and Development Program of China (2020YFA0309400).
K.J.R.\ acknowledges the support from the Shanghai Sailing Program (Grant No. 20YF1430600). G.Q.Y.\ thanks the National Natural Science Foundation of China (NSFC, Grant No.\ 11874409) and the Beijing Natural Science Foundation (Grant No.\ Z190009) for their financial support.
T.H.\ acknowledges support from the Engineering and Physical Sciences Research Council (UK) through grant EP/N032128/1.

\hfill


\noindent

\noindent {\footnotesize{{$^\ast$These authors contributed equally to the work.}} \\
\noindent {\footnotesize{Gerrit.vanderLaan@diamond.ac.uk }}\\
\noindent {\footnotesize{Thorsten.Hesjedal@physics.ox.ac.uk}}\\
\noindent {\footnotesize{duhf@hmfl.ac.cn}}\\
\noindent {\footnotesize{guoqiangyu@iphy.ac.cn}}\\
\noindent {\footnotesize{zhangshl1@shanghaitech.edu.cn}}


%

\end{document}